\documentstyle[seceq,preprint,eclepsf]{jpsj}

\def\runtitle{Classical Threshold Field }
\def\runauthor{Masanori {\sc Yumoto}, Hidetoshi {\sc Fukuyama} and Hiroshi {\sc Matsukawa}}

\title
{
Effect of Local Inhomogeneity on Nucleation; \\
Case of Charge Density Wave Depinning 
}

\author
{ 
Masanori {\sc Yumoto}\footnote{E-mail: yumoto@watson.phys.s.u-tokyo.ac.jp},
Hidetoshi {\sc Fukuyama} 
and Hiroshi {\sc Matsukawa}$^{1}$
}

\inst
{
Department of Physics, Tokyo University, Tokyo 113\\
$^1$Department of Physics, Osaka University, Toyonaka 560
}

\abst
{The spatial inhomogeneities are expected to affect nucleation process in an essential way. 
 These effects are studied theoretically by considering the case of the depinning 
 of the charge 
 density wave as a typical example. The threshold field of the depinning of 
 the one-dimensional commensurate charge density wave with one impurity has been 
 examined based on the phase Hamiltonian at absolute zero. 
 It is found that the threshold field is 
 lowered by a finite 
 amount compared to that in the absence of an impurity. }

\kword
{nucleation, inhomogeneity, frustration, charge density wave (CDW), commensurability, 
impurity, threshold field, phase Hamiltonian }

\begin{document}
\sloppy
\maketitle

\section{Introduction}
Nucleation  is one of the most drastic phenomenon in various fields of  
physics, chemistry, biology, and also in engineering.~\cite{rf:Hanggi}  
Especially, the nucleation in condensed matter physics is 
most interesting in the sense that it can be controlled by   
such parameters as pressure, temperature, electric and magnetic fields so on. 
In general nucleation is defined as a phenomenon where a new phase appears locally 
in space.

The theoretical analysis of nucleation was given by 
Langer,~\cite{rf:Langer} who investigated the problem of reversing of the direction of 
magnetization in a ferromagnetic system. 
In the process of reversing, changes of magnetization is found to be  
not uniform in space but it is triggered by 
the appearance of the magnetic bubbles called  
droplets. 
While this theory offers the fundamental understanding of the nucleation in a homogeneous medium, 
we know that a local inhomogeneity in actual 
systems plays 
important roles in the nucleation process, which will be studied in this paper. 
For explicit studies we choose the case of the charge density wave (CDW)~\cite{rf:1} 
depinning as a typical example, because 
the field theory which is described by the phase Hamiltonian has been established,~\cite{rf:HFHT}
and the concept of the nucleation can be defined clearly in this case as  
the appearance of the spatially local non-uniform structure 
of the phase variable which triggers the depinning.

CDW is one of the characteristic state of quasi-one-dimensional conductors, 
where translational symmetry is spontaneously broken. 
The gapless sliding mode~\cite{rf:LRA} that 
incarnates Goldstone mode can be pinned by the external objects 
which break the translational symmetry.  If the external object is 
the underlying lattice whose spatial periodicity is commensurate to CDW's, 
it is called commensurate CDW. In this case CDW behaves as an insulator. 
However, if electric field is applied,    
CDW starts to move above some critical field called 
threshold field.  To estimate the threshold field theoretically, 
the phase Hamiltonian approach~\cite{rf:2} is useful. 
With the phase Hamiltonian approach, the classical threshold field can be
derived as the field at which  
the potential barrier for the sliding disappears. 
In this paper we investigate classically  
the depinning processes of one-dimensional commensurate CDW 
with inhomogeneity at absolute zero. 

In \S2 we introduce our model based on the phase Hamiltonian. In \S3 we review briefly 
in a homogeneous case. Below the 
uniform depinning field, 
the nucleation requires a finite excitation energy. 
In \S4 we examine the ground state in the presence of an 
impurity and in \S5 we investigate the threshold field in this case. The threshold 
field can be smaller than the uniform depinning field and depinning sets at 
the impurity site. In \S 6 the potential curve of our model is shown. Our conclusion 
and discussion are given in \S7.  
The effects of three dimensionality and fluctuations, quantum and thermal, 
will be studied separately.

\section{The Model}
We investigate the one-dimensional commensurate CDW with 
one impurity located at $X_{i}$. The Lagrangian is then given by 
\begin{equation}
\cal L  \mit =  - \int {\rm d}X \left[  A\left(\frac{\partial\phi}{\partial X} \right)^2 
- F\phi + g \left(1 - \cos(M \phi) \right) 
- V_{i}\cos(2 k_{{\rm F}} X + \phi)\delta(X-X_{i})  \right] .
\end{equation} 
Here, the first term in the integration is the elastic energy,  
$A=(\hbar v_{{\rm F}})/(4 \pi)$, with 
$v_{{\rm F}}$ being the Fermi velocity.
The second term is the energy associated with 
the electric field, $F = ({\rm e}E)/(2 \pi)$, with  
$E$ and $- {\rm e}$ (${\rm e} > 0$) being electric field and the electron charge, respectively. 
The third term is the commensurability energy, $g=(| \Delta |^2 / \varepsilon_{{\rm F}}) 
n_{{\rm e}} (| \Delta |/W )^{M-2}$, $|\Delta|$ being the Peierls gap, and 
$\varepsilon_{{\rm F}}$,  
$n_{{\rm e}}$, $W$ are the Fermi energy, the density of electron, and the width of 
the original band, respectively.  
Here $M$ is the degree of commensurability, $M =\pi/(k_{{\rm F}} a)$, 
$k_{{\rm F}}$ is the Fermi wave number,  
$a$ is the lattice spacing.
The last term is the coupling energy of CDW to one impurity.
In this Lagrangian, there exists a characteristic length, 
$\xi = \sqrt{(2 A)/(M^2 g) }$, which 
is the phase coherence length due to the commensurability. 
We make the Lagrangian dimensionless through scaling by this $\xi$; i.e.
$x =X/\xi$,  
$\cal L \mit = M^2 g \xi L$, 
where $L$ is given by  
\begin{equation}
L = - \int {\rm d}x \left[  \frac{1}{2}\left(\frac{\partial\phi}{\partial x} \right)^2 
- \varepsilon\phi + \frac{1}{M^2}\left(1 - \cos(M \phi) \right)
- v\cos(\chi + \phi)\delta(x-x_{i}) \right]. \label{eq:2.4}
\end{equation}
Here  
$\varepsilon = F/(g M^2)$,  
$v = V_{i}/(g \xi M^2)$. In this paper, we assume $\varepsilon \geq 0 $. 
Further, we define $x_{i}=X_{i}/\xi$ and 
$\chi= (2 \pi z)/M $, where  
$z=X_{i}/a $ characterizes the location of an impurity relative to the site where the 
energy gain by commensurability is maximum.  
The range of $\chi$ is $-\pi/M \leq \chi \leq \pi/M$, 
because the Lagrangian has the periodicity of $2\pi/M$ with respect to the phase.

\section{The Homogeneous Case}
First of all, we examine the case of $v=0$, namely the homogeneous case.~\cite{rf:3}

\begin{figure}
\epsfigure{file=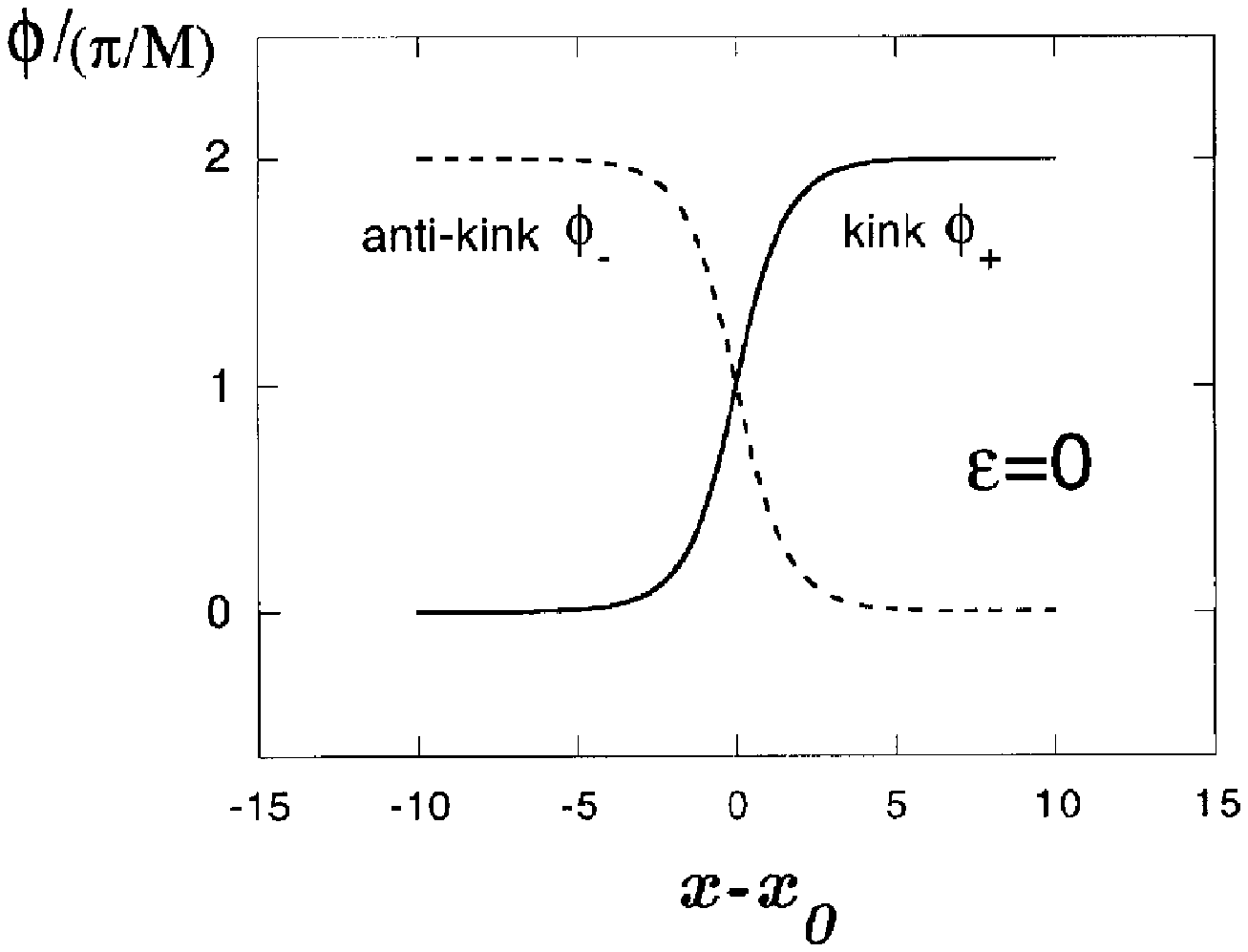,height=6cm}
\caption{The classical solution in the case of $\varepsilon = 0$. The solid line is a kink 
and the dotted line is an anti-kink. }
\label{fig:kink}
\end{figure}

The stable configuration of the phase is determined by 
varying Lagrangian;
\begin{equation}
- \phi'' - \varepsilon + \frac{1}{M}\sin(M \phi)=0. \label{eq:3.1} 
\end{equation}
When $\varepsilon = 0$, eq.~(\ref{eq:3.1}) is the sine-Gordon equation, 
and has two kinds of solutions, which are a trivial one, 
$\phi=0$,   
and the kink solution, which is given by 
\begin{equation}
\phi(x)= \frac{4}{M}\arctan\left[\exp(\pm(x-x_{0})\right] \equiv \phi_{\pm}(x-x_{0}), 
\label{eq:3.3}  
\end{equation} 
and shown in Fig.~\ref{fig:kink}.
Here $x_{0}$ is the center of the kink. 
Among these solutions, the lowest energy solution, namely, the ground state solution is 
$\phi=0$, and the kink and the anti-kink are excitations with 
finite excitation energy.

\begin{figure}
\epsfigure{file=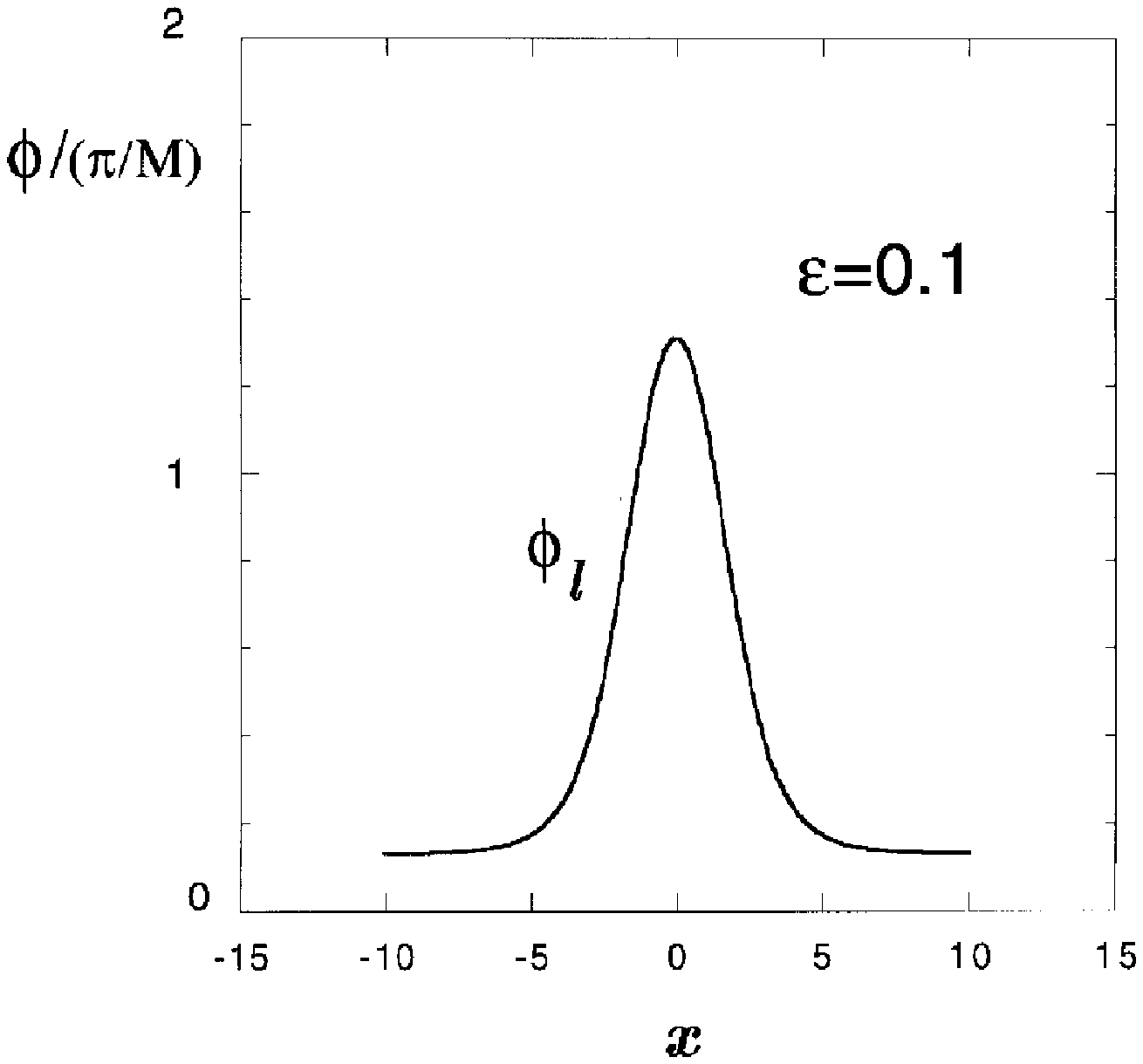,height=6cm}
\caption{The non-uniform solution, $\phi_{l}(x)$, of the classical equation in the case of 
$\varepsilon=0.1$.  }
\label{fig:soliton}
\end{figure}

In the case of $\varepsilon \neq 0$ the equation of the classical configuration  has 
also an uniform solution,   
\begin{equation}
\phi =\frac{1}{M}\arcsin(\varepsilon M),  
\end{equation} 
and the non-uniform solution with local spatial variation, $\phi_{l}(x)$, 
which is given by 
\begin{equation}
\phi_{l}' = -\left[ 2 \left[ \varepsilon \left( \frac{1}{M} \arcsin(\varepsilon M) - 
\phi_{l} \right)
+ \frac{1}{M^2}\left(\sqrt{1-(\varepsilon M)^2} 
- \cos(M \phi_{l})\right)\right]\right]^{\frac{1}{2}}. \label{eq:3.5}
\end{equation}
This equation is derived by the integration of eq.~(\ref{eq:3.1}) with the boundary condition, 
$\phi(\infty)=(1/M)\arcsin(\varepsilon M)$ and 
$\phi'(\infty)=0$.
The solution, $\phi_{l}(x)$, is shown in Fig.~\ref{fig:soliton}. 
We notice that the non-uniform solution, $\phi_{l}(x)$, is kink pair solution.    

\begin{figure}
\epsfigure{file=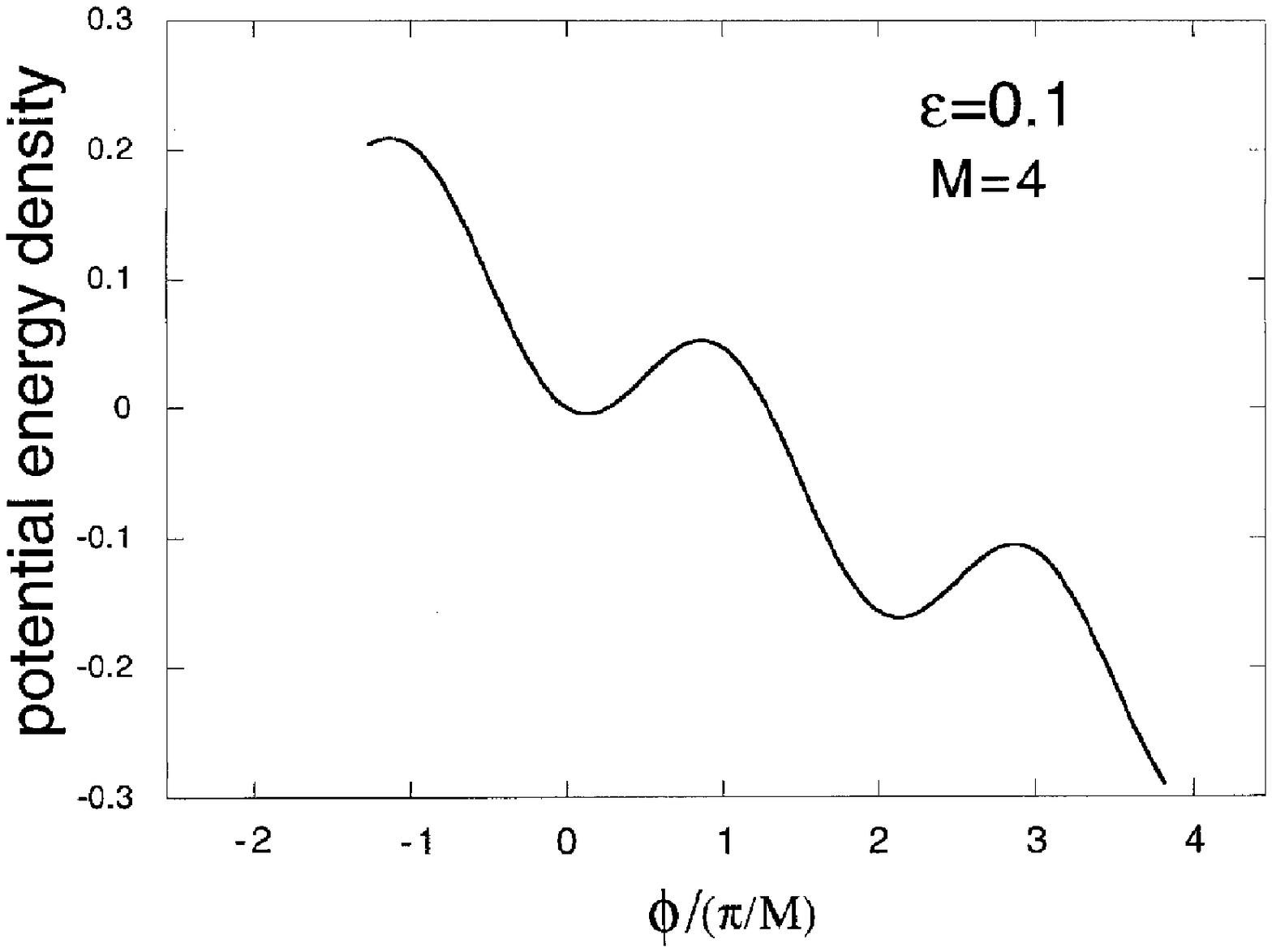,height=6cm}
\caption{The potential for the uniform phase at $\varepsilon=0.1$ for a choice of $M=4$. }
\label{fig:1}
\end{figure}

In the presence of the electric field, the cosine-type potential for the uniform phase  
is tilted as is shown in Fig.~\ref{fig:1}. The potential barrier disappears 
at $\varepsilon=1/M \equiv \varepsilon_{T}$, which is the classical depinning field of 
the uniform CDW.  

On the other hand, 
if the kink pair is excited, CDW will also depin. This is another type of depinning process 
which sets locally in space and considered to be the nucleation process.
However, below the threshold field, $\varepsilon_{T}$, of the uniform depinning,  
the kink pair is always the excitation, which requires a finite excitation energy.
Therefore the nucleation does not take place in a homogeneous case without 
quantum nor thermal fluctuations.

\section{Ground State in the Presence of an Impurity}    
Next, we consider a case with an impurity, $v \neq 0$.
The stable configuration of the phase is now given by  
\begin{equation}
- \phi'' - \varepsilon + \frac{1}{M}\sin(M\phi) 
+ v\sin(\chi + \phi)\delta(x-x_{i}) =0,  \label{class}
\end{equation}
with   
the boundary conditions 
\begin{equation}
\phi(\pm \infty)=\frac{1}{M}\arcsin(\varepsilon M) \; , \;\;
\phi'(\pm \infty)=0.  \label{bouncla}
\end{equation}

\begin{figure}
\epsfigure{file=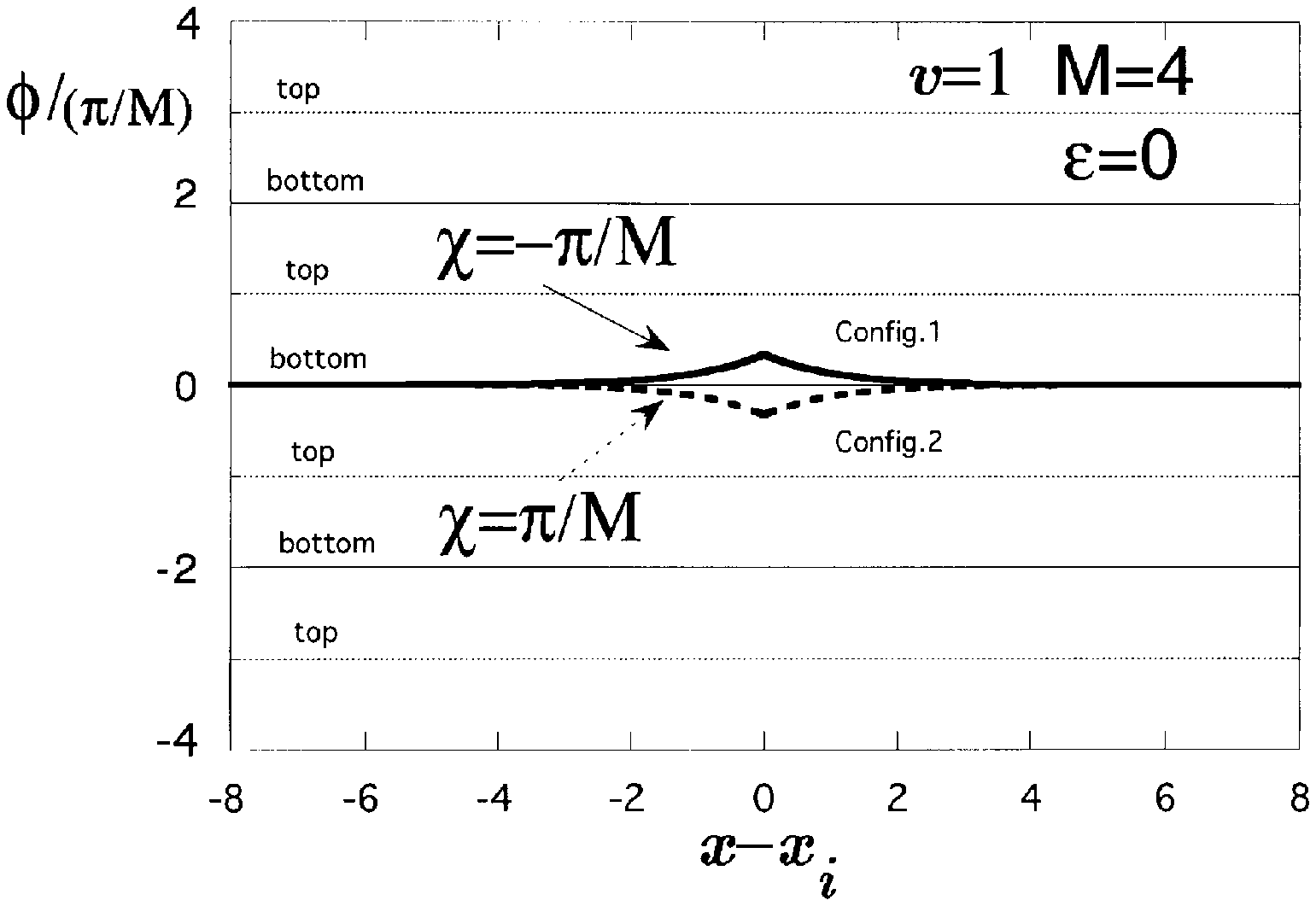,height=6cm}
\caption{The two kinds of the configurations 
with respect to $\chi$. This figure is for a choice of $M=4$, $v=1$ and $\varepsilon=0$.  }
\label{fig:conf}
\end{figure}

In the case of $\varepsilon = 0$, the solution is given analytically. 
Depending on the range of $\chi$, there are two kinds of configurations. 
If $\chi$ is in the range of 
$-\pi/M \leq \chi \leq 0$, the solution is located in the region between 
$0$ and $\pi/M$ (Config.1)  
as shown in Fig.~\ref{fig:conf}.  
If $\chi$ is in the range of $0 \leq \chi \leq \pi/M$, the solution 
is located in the region between $-\pi/M$ and $0$ (Config.2). 
We take electric field $\varepsilon \geq 0$. Under this condition, Config.1   
has more tendency to depin than Config.2. Namely, in the case of $-\pi/M \leq \chi \leq 0$, 
the threshold field can be smaller than that of uniform depinning as is disclosed in \S5; 
in the case of $0 \leq \chi \leq \pi/M$, however, the threshold field is same as that of 
uniform depinning. Therefore, we consider only the case of $-\pi/M \leq \chi \leq 0$. This 
choice is justified by the discussion in \S7. 

\begin{figure}
\epsfigure{file=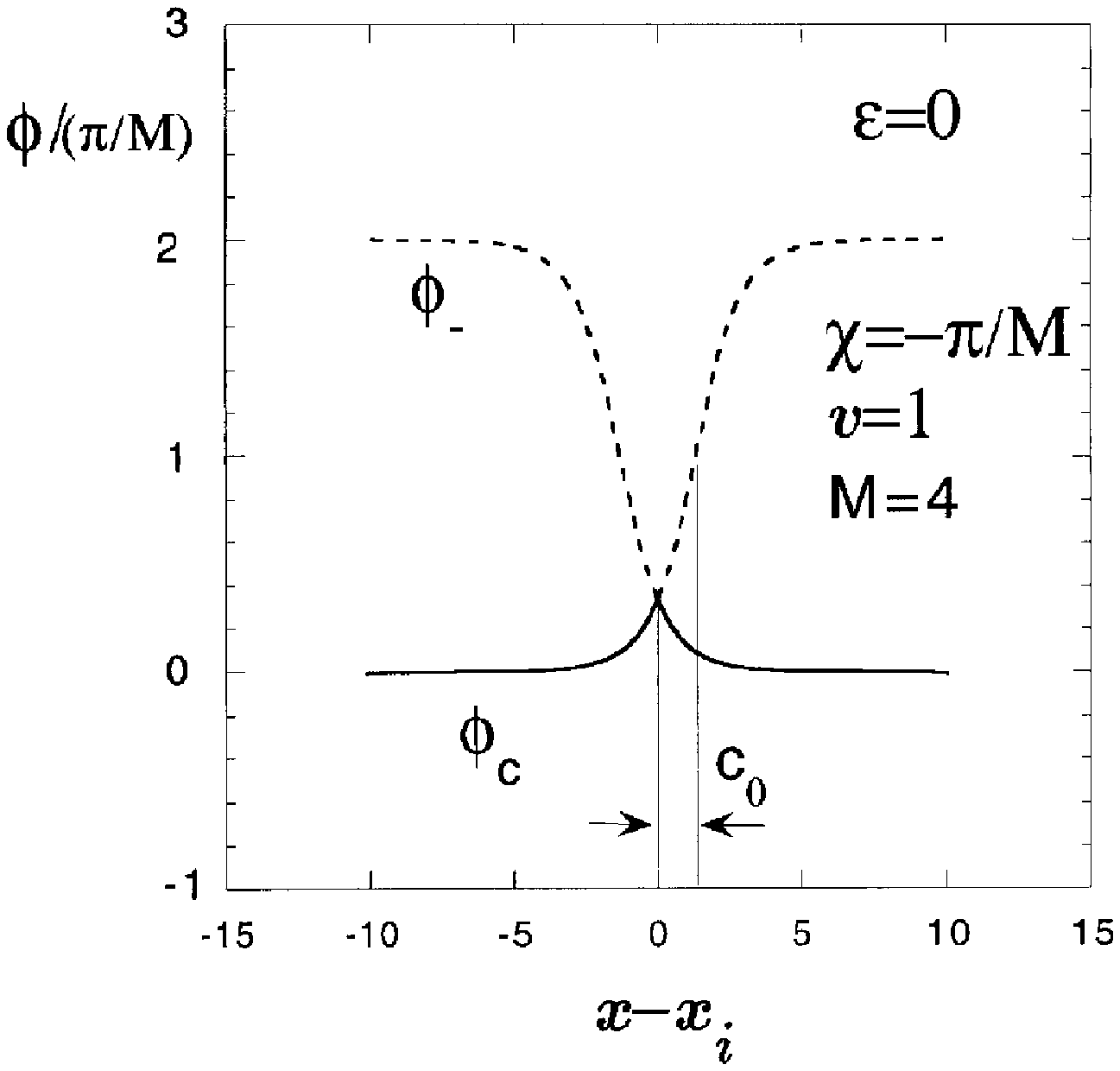,height=6cm}
\caption{ The solid line is the stable solution with an impurity 
at $x=x_{i}$ for a choice of $M=4$, $v=1$, $\chi=-\pi/M$ and $\varepsilon = 0$. }
\label{fig:e0}
\end{figure}

The solution of Config.1 is obtained in terms of $\phi_{-}(x)$ given in eq.~(\ref{eq:3.3}), 
\begin{eqnarray}
\phi_{c}(x) &=& \frac{4}{M}\arctan\left[\exp\left(-|x-x_{i}|-c_{0}\right)\right]  
\label{clasol} \\
             &=& \phi_{-}(|x-x_{i}|+ c_{0}),  
\end{eqnarray}
where $c_{0}$ is the parameter which is determined by the equation, 
\begin{equation}
\frac{4}{M}\frac{1}{\cosh(c_{0})} + v \sin \left(\chi + \frac{4}{M}\arctan[\exp(- c_{0})]
\right) = 0.      \label{connection}
\end{equation}
This equation is due to the requirement that the jump of 
the first derivative of $\phi_{c}(x)$ at 
$x_{i}$ should match the third term in the left hand side of eq.~(\ref{class}).
The solution $\phi_{c}(x)$ is considered to be a kink and an anti-kink connected 
at the impurity site as shown in Fig.~\ref{fig:e0}.  
Further the parameter $c_{0}$ is equivalent to the distance from the impurity site to the 
center of the kink as indicated in Fig.~\ref{fig:e0}. 

\begin{figure} 
\epsfigure{file=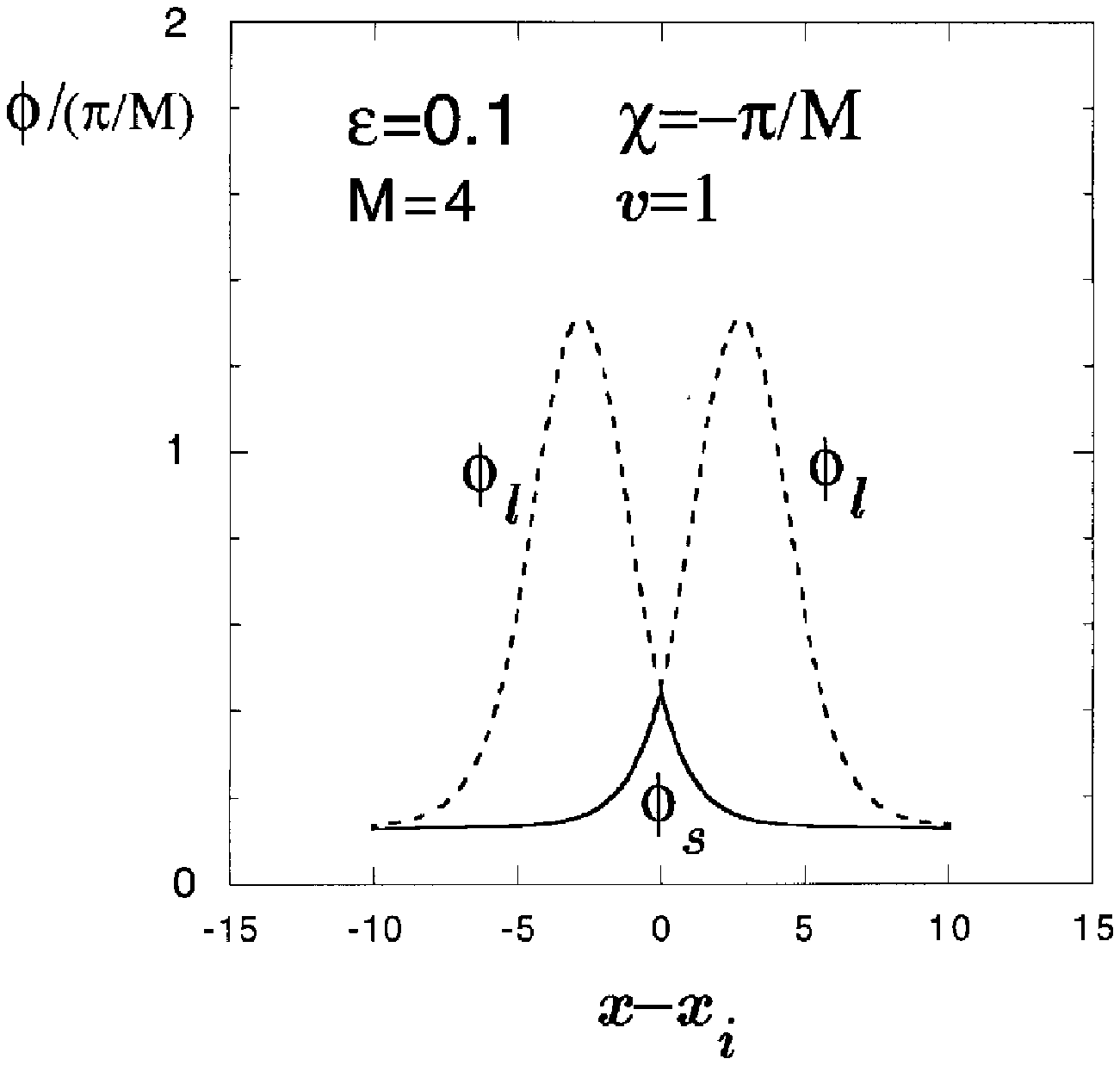,height=6cm}
\caption{The solid line is the stable solution with an impurity for a choice of  
 $M=4$, $v=1$, $\chi=-\pi/M$ and $\varepsilon = 0.1$. }
\label{fig:2}
\end{figure}

In the case of $\varepsilon \neq 0$, the stable solution, $\phi_{s}(x)$, 
in the presence of an impurity 
given by eq.~(\ref{class}) is obtained by the connection 
at the impurity site  
of two non-uniform solutions, one of which, $\phi_{l}(x)$, has been shown in Fig.~\ref{fig:soliton}.  
An example of the stable solution, $\phi_{s}(x)$, is shown in Fig.~\ref{fig:2}. 

Though the stable solution, $\phi_{s}(x)$, is 
obtained only numerically, the information of the phase value at 
the impurity site is obtained.
The stable solution, $\phi_{s}(x)$, is given as follows; 
\begin{equation}
\phi_{s}(x) = \phi_{l}(|x-x_{i}| + c)   \label{classsol}
\end{equation}
where $\phi_{l}(x)$ is the non-uniform solution of eq.~(\ref{eq:3.1}) and $c$ is 
a parameter to be determined to satisfy the boundary 
condition given by eq.~(\ref{bouncla}).
When we consider $\phi_{s}(x)$, it is necessary to consider only the range of solid line 
in Fig.~\ref{fig:2}.

Substituting eq.~(\ref{classsol}) into eq.~(\ref{class}), we obtain  
\begin{equation}
- 2 \phi_{l}'(c) + v \sin(\chi + \phi_{l}(c)) = 0. \label{cla1}
\end{equation}
By use of eq.~(\ref{eq:3.5}), eq.~(\ref{cla1}) is rewritten as  
\begin{equation}
\left[ 2 \left[ \varepsilon \left( \frac{1}{M} \arcsin(\varepsilon M) - \phi_{l}(c) \right)
+ \frac{1}{M^2}\left(\sqrt{1-(\varepsilon M)^2} 
- \cos(M \phi{l}_(c))\right)\right]\right]^{\frac{1}{2}}
+ \frac{1}{2}v\sin(\chi + \phi_{l}(c)) = 0.    \label{result}
\end{equation}
This equation determines the phase value, $\phi_{l}(c)$, (instead of $c$) 
at the impurity site of 
the stable solution when 
the electric field, $\varepsilon$, is given. 
It means that instead of the boundary condition at infinity, 
eq.~(\ref{bouncla}), we  obtain a boundary condition at the impurity site.

\section{Threshold Field in the Presence of an Impurity} 
The depinning threshold field, $\varepsilon_{c}$, is   
determined as the field at which the stable solution becomes unstable.~\cite{rf:4} 
The instability of the solution is triggered
by the onset of a zero eigenvalue of the fluctuation mode  
around the stable solution.~\cite{rf:5}

The eigenvalue equation of the fluctuations, $\delta \phi(x)$, 
around the stable solution, $\phi_{s}(x)$, is 
the second variational equation 
of the Lagrangian; i.e.
\begin{equation}
\left[ - \frac{\partial^2}{\partial x^2} + \cos(M \phi_{s}) 
+ v\cos(\chi + \phi_{s})\delta(x-x_{i}) \right]\delta \phi = \lambda \delta \phi,  
\label{fluc}
\end{equation}
where $\lambda$ is the eigenvalue. 
Hence we consider the case of $\lambda=0$ to determine the threshold field, $\varepsilon_{c}$.

First, we consider the fluctuation around the solution, $\phi_{l}(x)$, in the case of $v=0$. 
The equation of the zero-mode fluctuation, $\delta\phi_{l}(x)$, around $\phi_{l}(x)$ is 
\begin{equation}
\left[ - \frac{\partial^2}{\partial x^2} + \cos(M \phi_{l}) \right] \delta\phi_{l} 
= 0
\end{equation}
and the boundary condition is 
\begin{equation}
\delta\phi_{l}(\infty) = 0 \; , \;\; \delta\phi_{l}'(\infty) = 0.
\end{equation}
Its solution is 
\begin{equation}
\delta\phi_{l}(x) \propto \phi_{l}'(x).
\end{equation}

Next, we consider the case of $v \neq 0$. The threshold field, $\varepsilon_{c}$, is 
determined as a particular value of  
the electric field at which   
zero-mode fluctuations in both sides of the impurity are connected.  
The zero-mode fluctuation around the stable solution in each region must be expressed as 
\begin{eqnarray}
\delta\phi(x) &\propto& - ( 2 \theta(x-x_{i}) - 1) \phi_{s}'(x)  \\
                 &=& - \phi_{l}'(|x-x_{i}|+c),  \label{flu1}
\end{eqnarray}
by noting
\begin{equation}
\phi_{s}'(x) = (2\theta(x-x_{i})-1)\phi_{l}'(x),
\end{equation}
where $\theta(x)$ is the step function. 
Substituting eq.~(\ref{flu1}) into eq.~(\ref{fluc}), 
we obtain 
\begin{equation}
2\phi_{l}''(c) - v\cos(\chi + \phi_{l}(c))\phi_{l}'(c) = 0 .   \label{conflu}
\end{equation}
At the threshold field, $\varepsilon_{c}$, eq.~(\ref{conflu}) should be satisfied. 
By use of eqs.~(\ref{eq:3.1}) and (\ref{cla1}), eq.~(\ref{conflu}) is expressed as follows; 
\begin{equation}
2 \left( -\varepsilon + \frac{1}{M}\sin(M \phi_{l}(c)) \right) 
- \frac{1}{4} v^2 \sin \left[ 2(\chi + \phi_{l}(c)) \right] =0. \label{5.9}
\end{equation}
Hence the threshold field, $\varepsilon_{c}$, is determined as a value of $\varepsilon$ 
at which both eqs.~(\ref{result}) and (\ref{5.9}) have a common solution, $\phi_{l}(c)$, 
for each fixed values of $\chi$ and $v$. 
 
\begin{figure}
\epsfigure{file=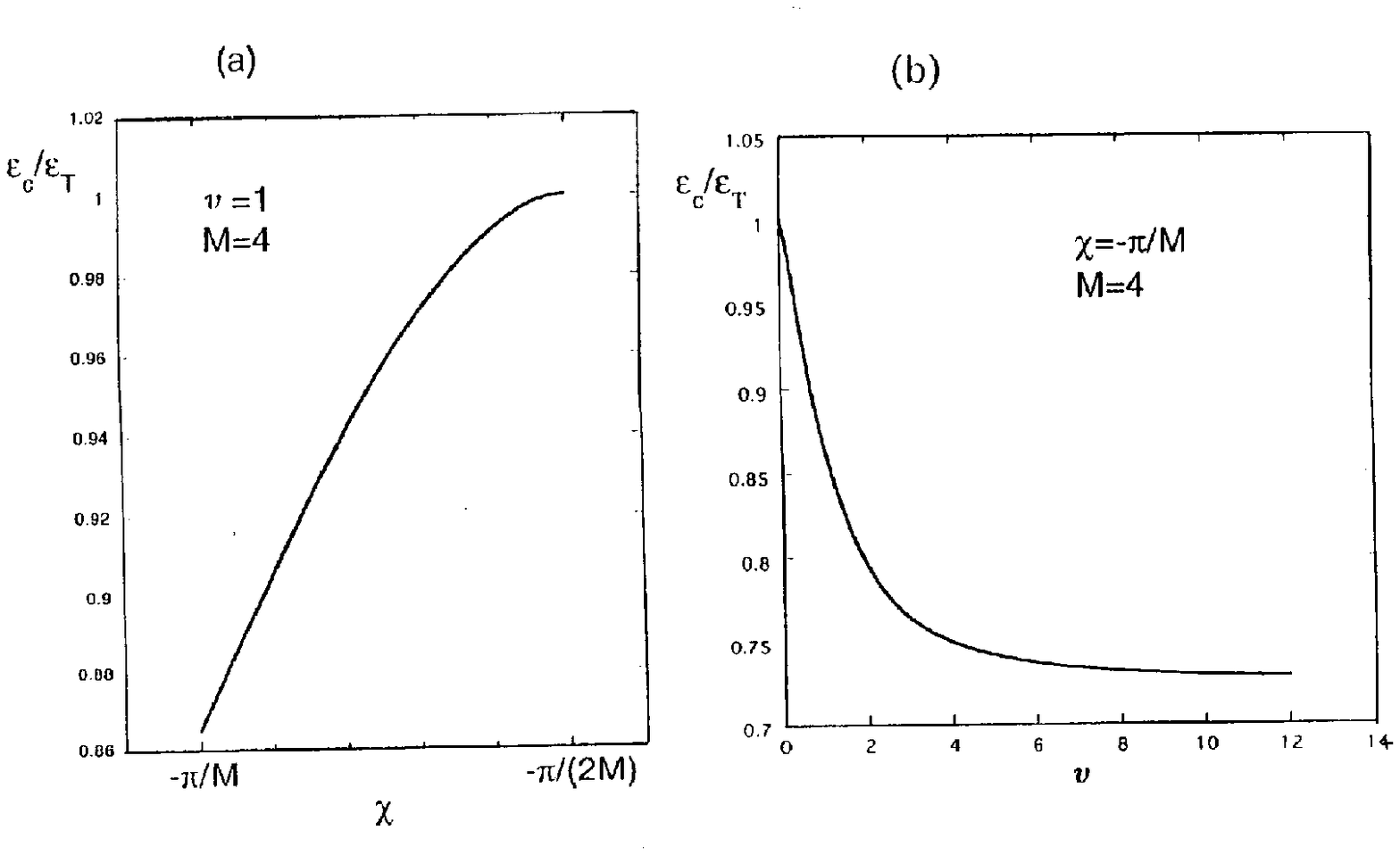,height=8cm}
\caption{The threshold field in the presence of an impurity,  
normalized by that of the uniform case, for a choice of $M=4$. 
The threshold field for a choice of $v=1$ is 
shown in (a) and for a choice of $\chi=-\pi/M$ is shown in (b). }
\label{fig:3}
\end{figure}

The solutions of eqs.~(\ref{result}) and (\ref{5.9}) for $M=4$ are shown in Fig.~\ref{fig:3}. 
The threshold field, $\varepsilon_{c}$, is normalized by the threshold field, $\varepsilon_{T}$,
of the uniform case. 
When $v$ is fixed, the normalized threshold field goes to 1 as $\chi$ tends to $-\pi/(2M)$. 
It is noted that the threshold field goes to a finite value, $\varepsilon_{\infty}$,  
even if $v \rightarrow \infty$. This finite 
value, $\varepsilon_{\infty}$, is given by the equation, 
\begin{equation}
\varepsilon_{\infty} (\frac{1}{M}\arcsin(\varepsilon_{\infty} M) + \chi ) 
+ \frac{1}{M^2} (\sqrt{1-(\varepsilon_{\infty} M)^2}-\cos(M \chi)) = 0 .
\end{equation}
Through a numerical calculation, we find that for a choice of $\chi = -\pi/M$ and $M=4$, 
the normalized threshold field tends to 
$\varepsilon_{\infty}/\varepsilon_{T} \simeq 0.725$ as $v$ tends to infinity.

The reason why $\chi =- \pi/(2M) $  is critical is that  $\chi = - \pi/(2M) $ is 
the value which determines whether the commensurability potential and the 
impurity potential compete or not at the impurity site, namely, the system contains 
frustration or not. To investigate the frustration, we consider  
the local potential energy $U_{i}$ at the impurity site,   
\begin{equation} 
U_{i}(\phi(x_{i})) = - \frac{1}{M^2}\cos(M\phi(x_{i})) - v\cos(\chi + 
\phi(x_{i})). 
\end{equation}
The first term of the right hand is the commensurability potential and the second term is 
the impurity potential. If the potential energy by the impurity is minimized, 
that is $\chi + \phi(x_{i}) = 0$, 
$U_{i}(\phi(x_{i}))$ is 
\begin{equation}
U_{i}(-\chi) = - \frac{1}{M^2}\cos(M\chi) - v.
\end{equation}
In the case of $- \pi/M \leq \chi < - \pi/(2M) $, the commensurability potential and 
the impurity potential compete and the system contains frustration. 
In the case of $ - \pi/(2M) < \chi \leq 0$, however, they do not compete 
and the system contains no frustration. 
When the system is frustrated the effect of the impurity potential to the threshold field 
is important. However, if the system is not frustrated, the impurity 
potential is irrelevant.

\begin{figure}
\epsfigure{file=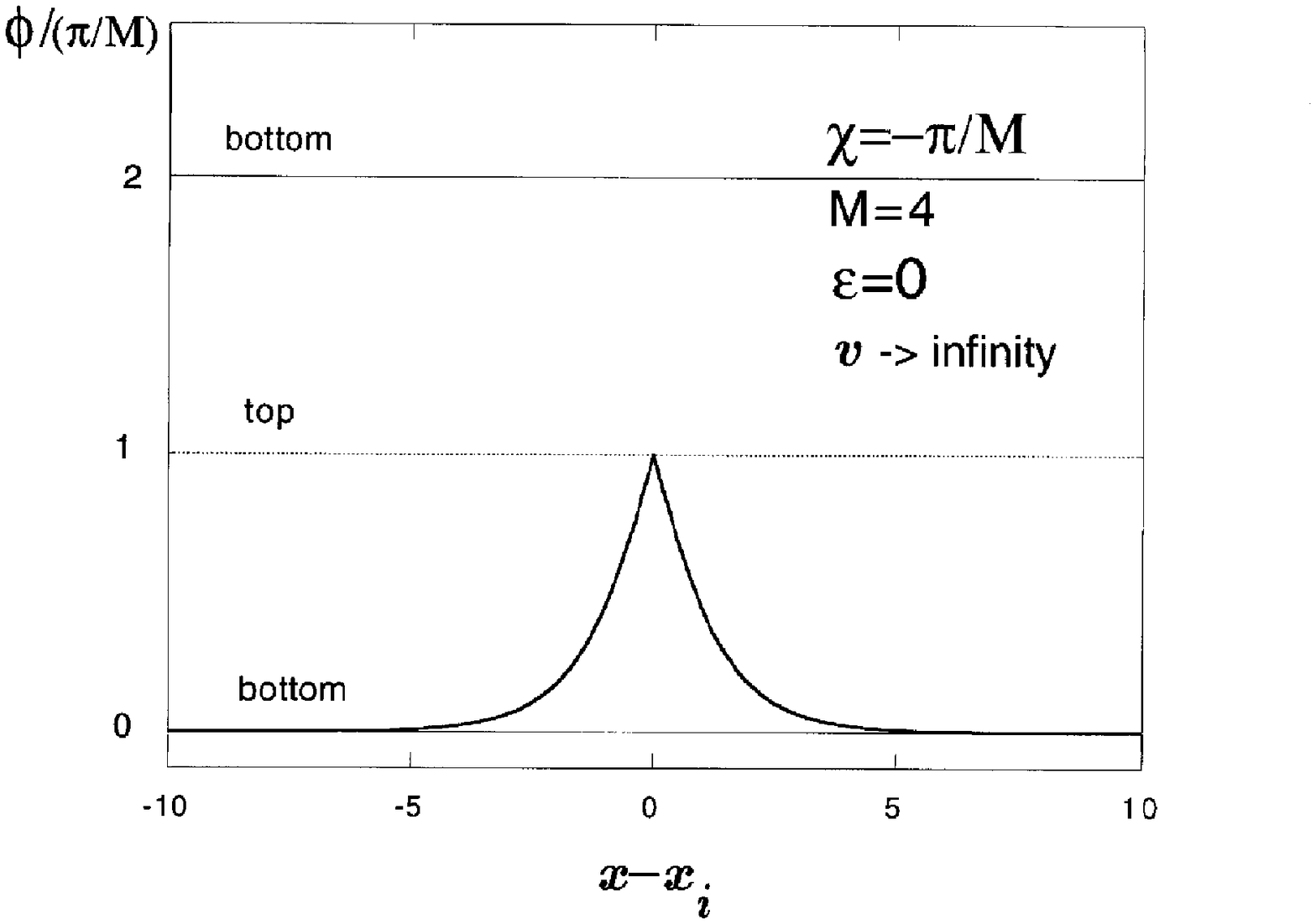,height=6cm}
\caption{ The classical solution for a choice of $M=4$, $\chi=-\pi/M$, 
$\varepsilon=0$ and $v \rightarrow \infty$. }
\label{fig:5}
\end{figure}

The reason why the threshold field is finite even if $v$ tends to infinity 
is that the depinning object is not a particle, but a string. 
In the limit of $v \rightarrow \infty$ with $\chi = -\pi/M $ which is the optimal case, 
the cusp of the ground state configuration of the phase is on the top of the barrier of  
the uniform potential (Fig.~\ref{fig:5}). 
The cusp, 
however, is just a part of the string, and a finite field is necessary for a string to go 
over the uniform potential barrier.

\section{The Potential Curve in the Presence of an Impurity}
\begin{figure}
\epsfigure{file=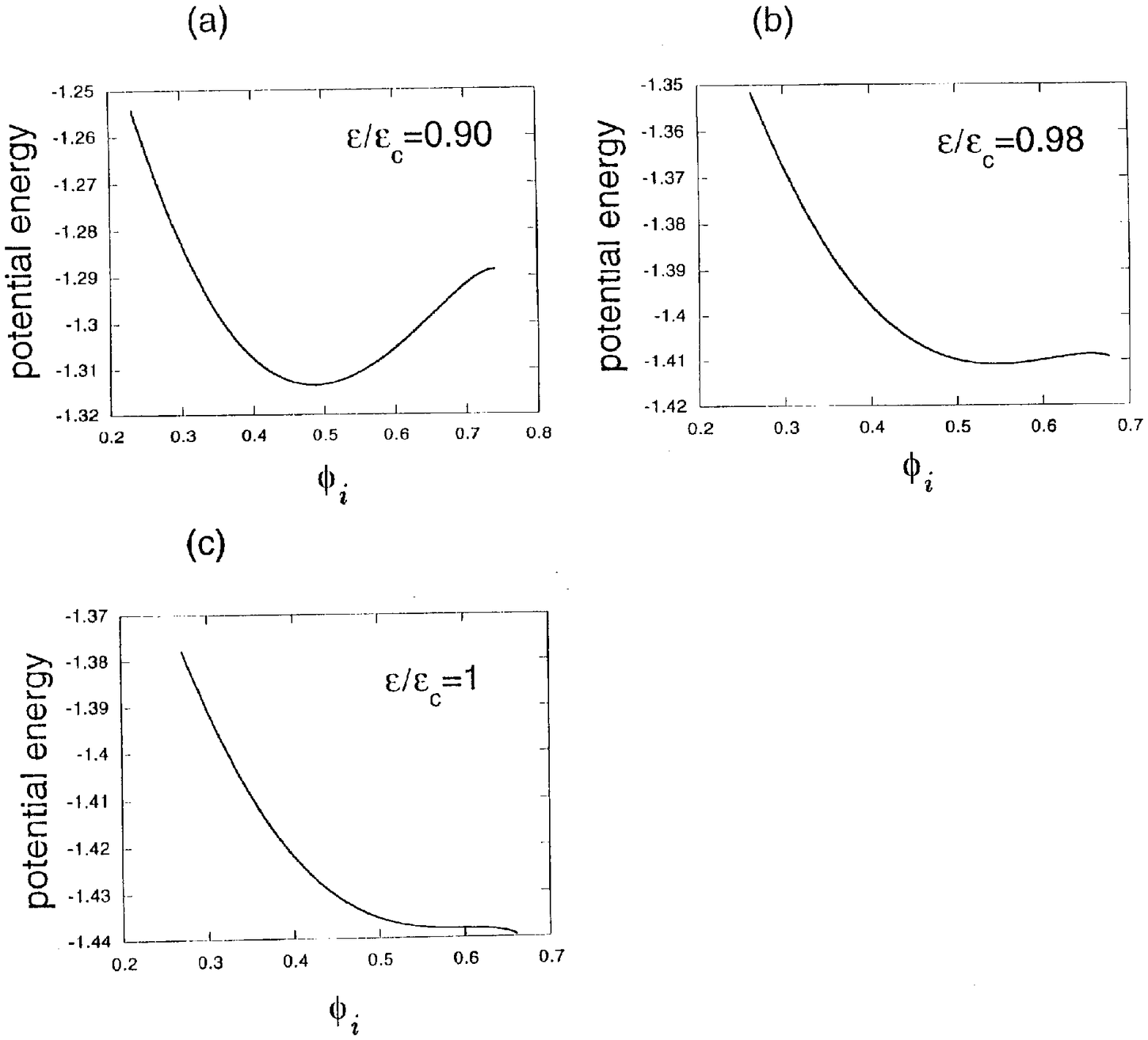,height=12cm}
\caption{ The potential curve in the presence of an impurity for a choice of 
 $M=4$, $\chi=-\pi/M$ and $v=1$; 
(a) is in the case of $\varepsilon/\varepsilon_{c}=0.90$, 
(b) is in the case of $\varepsilon/\varepsilon_{c}=0.98$ and 
(c) is in the case of $\varepsilon/\varepsilon_{c}=1$.}
\label{fig:curve}
\end{figure}

To clarify the meaning of $\varepsilon_{c}$, we consider a potential curve in the 
presence of an impurity. In a homogeneous case the phase is uniform in 
the ground state, and hence 
the potential energy density (Fig.~\ref{fig:1}) can be expressed 
with respect to the uniform phase variable. 
With an impurity, however, the 
phase is not uniform, and therefore we should consider the potential curve in a functional 
space. 
In this case the most suitable variable is 
the phase value, $\phi_{i}$, at the impurity site  
with $\chi$, $v$ and $\varepsilon$ fixed. For each value of $\phi_{i}$, we can determine the 
lowest energy configuration; we solve eq.~(\ref{eq:3.1}) under the boundary condition of 
$\phi(x_{i})=\phi_{i}$ and 
\begin{equation}
\phi'(x_{i}) = -\left[ 2 \left[ \varepsilon \left( \frac{1}{M} \arcsin(\varepsilon M) - 
\phi_{i} \right)
+ \frac{1}{M^2}\left(\sqrt{1-(\varepsilon M)^2} 
- \cos(M \phi_{i})\right)\right]\right]^{\frac{1}{2}},  
\end{equation}
which is similar to eq.~(\ref{eq:3.5}). 
Substituting this optimal configuration into eq.~(\ref{eq:2.4}), we determine its 
energy and obtain the potential curve as 
the function of $\phi_{i}$ which is shown 
in Fig.~\ref{fig:curve}. 
While there exists an energy barrier below the threshold field, $\varepsilon_{c}$, 
(Fig.~\ref{fig:curve} (a) and (b)),  
the barrier disappears at $\varepsilon_{c}$ (Fig.~\ref{fig:curve} (c)) and CDW depins. 

\section{Conclusion and Discussion}
We conclude that the existence of an impurity in the model one-dimensional commensurate CDW  
causes the lowering of 
threshold fields by a finite amount due to the appearance of the local 
change of the phase variable, which is considered as nucleation, near an impurity site. 
Below the threshold field, 
$\varepsilon_{T}$, of the 
uniform depinning, such local change in the homogeneous case 
always requires a finite excitation energy.  However, the nucleation 
in the presence of an impurity can take place without any excitation energy 
at field, $\varepsilon_{c}$, which is smaller than $\varepsilon_{T}$. 
The main reason for the lowering of the threshold field is due to the frustration which is 
caused by the competition between the commensurability potential and the impurity potential.

Our result will be applied to the commensurate CDW with dilute but 
with macroscopic number of impurities where 
the inverse of the impurity density is smaller than the 
phase coherence length, $\xi$.   
In such a case these will be a distribution of parameter $\chi$, and then 
that of the depinning field.  
However the depinning will be triggered by the nucleation at the optimal impurity site, 
$\chi = -\pi/M$.   

In actual experiments three dimensionality and quantum fluctuations~\cite{rf:matsu} 
will play important roles at low temperatures, which will be studied elsewhere.

To the best of our knowledge, our investigation is the first  
microscopic theoretical studies on nucleation triggered by the  
local inhomogeneity in the bulk region.

\section*{Acknowledgments}
We would like to thank Hiroshi Kohno for helpful discussions. 
One of us (M.Y.) is grateful for the kind hospitality received at 
Condensed Matter Physics Group of Osaka University.

\end{document}